# ALEXANDER BRUCE, SCOTLAND'S ACCIDENTAL 'SCIENTIFIC REVOLUTIONARY'

**Richard de Grijs**
*Department of Physics and Astronomy, Macquarie University,
Balaclava Road, Sydney, NSW 2109, Australia*
Email: richard.de-grijs@mq.edu.au

**Abstract:** The mid-17th century saw unprecedented scientific progress. With the Middle Ages well and truly over, the Scientific Revolution had begun. However, scientific advancement does not always proceed along well-planned trajectories. Chance encounters and sheer luck have important roles to play, although more so in the 17th century than today. In this context, the Scottish businessman and erstwhile royalist exile, Alexander Bruce (1629–1680), found himself in the right place at the right time to contribute significant innovations to the nascent pendulum clock design championed by contemporary natural philosophers such as Christiaan Huygens, Robert Moray, and Robert Hooke as the solution to the perennial 'longitude problem.' Bruce's fledgling interests in science and engineering were greatly boosted by his association with the brightest minds of the newly established Royal Society of London. From an underdog position, his innovations soon outdid the achievements of the era's celebrated scholars, enabling him to conduct some of the first promising sea trials of viable marine timekeepers. International collaboration became international rivalry as time went on, with little known Scottish inventions soon becoming part of mainstream clock designs.

## 1 TROUBLED TIMES

The history of science is replete with unexpected twists and turns, with major progress occasionally attributed to unlikely proponents. One would be hard-pressed to find a better example of such an unlikely proponent than one Alexander Bruce (1629–1680) of Broomhall, Fife (Scotland). Through a series of accidental developments and Bruce's fortuitous presence at the right time and in the right place, one of the Scientific Revolution's most pressing scientific problems—the need to determine one's longitude at sea, out of view of well-known shores—saw a number of unanticipated leaps forward when they were perhaps least expected.

Latitude determination is fairly straightforward. It determines one's position with respect to the Equator, so that determination is required of the height of the Sun at local noon, transiting the local meridian—the great circle that passes through the celestial North and South Poles and the observer's zenith—or (in the northern hemisphere) that of Polaris, presently the North Star. However, zero longitude can be defined arbitrarily, a property that has caused numerous political complications (de Grijs, 2017, Epilogue). The ancient Greek astronomer Hipparchus of Nicea (190–120 BCE) was already convinced that accurate, internally consistent maps should be based on astronomical measurements of latitudes and longitudes, and on triangulation. He proposed that one could determine positions East and West of his reference meridian by comparing the local time to an 'absolute' time, referenced at his zero meridian. Having established a reference meridian, one could then determine relative longitudes by taking into account that the Earth completes a full, 360° rotation in 24 hours. One hour therefore corresponds to 15° in longitude. Hipparchus was the first to realise that one's longitude can potentially be determined by means of accurate timekeeping, a realisation that perplexed numerous scholars in centuries to come. The intractable 'longitude problem' is an eminent example of a practical scientific pursuit of global importance. However, Bruce's early Scottish contribution to the problem's eventual resolution is less well-known.

> Another person who made a good figure … was Bruce, afterwards [second] Earl of Kincardine. He was both the wisest and the wittiest man that belonged to his country, and fit for governing any affairs but his own, which he, by a wrong turn and a love for the public, neglected to his ruin; for they, consisting much in works, coal, salt, and mines, required much care; and he was very capable of it, having gone far in mathematics. His thoughts went slow, and his words came much slower; but a deep judgment appearing in what he said and did, made a compensation. He had a noble zeal for justice, in which even friendship could not bias him, and a powerful sense of religion and virtue, which showed themselves with great lustre upon all occasions; and from such principles it is less wonder that he became a faithful friend and a merciful enemy … I may perhaps be inclined to carry his character too far, for he was the first man that entered into friendship with me. We continued for sixteen years in so entire a friendship that there was never either reserve or mistake between us all the while till his death. (Bishop Gilbert Burnet, family friend; Burnet, 1724–1734)

**1.1 Early life**

Little is known of Bruce's early life. It appears that by the 1650s, Alexander Bruce and his cousin William were trading from Swedish-controlled Bremen[1] (1657–1658) (Schuchard, 2002) and Hamburg (northern Germany) as well as from Rotterdam (1567–1660), where they established themselves among a community of English and (mostly) Scottish merchants and represented their family's commercial interests. Once the Bruces and their partner in trade, John Hamilton of Grange (1600–1674), had acquired a vessel, *de Weijenboem*, they set up a trade network from the continental German hinterland to the waterways connecting the Low Countries, France, and the British Isles. They transported timber from Norway and salt and coal from their estate in Scotland to La Rochelle, France, meanwhile supplying wine from France to the Dutch Republic. The profitability of their business remains questionable, however, since we have surviving correspondence implying that William Bruce seriously considered getting involved in commercial trading ventures to the West Indies or whaling in Greenland instead (Colvin, 1995), a trade they were already involved in by 1657. As we learn from correspondence (dated 6 October 1657; I will use Old Style/Julian calendar notation throughout, as used in Britain at the time) from the courtier and philosopher, Sir Robert, Earl of Moray (Murray)[2] (1608/9–1673) to Alexander Bruce,

> I send you this from Will … in short, his voyage and pains have made him no gains, but diminished his [wine and timber] stock very much [but] … he takes it most virtuously, to my great joy.[3]

And so from a merchant background, which was perhaps not altogether successful, the circumstances for Bruce to eventually play a decisive role in solving the key scientific problem of his time were tenuous at best. Nevertheless, through a series of what initially appeared to be unfortunate personal events, some of the earliest highly promising sea trials of a viable marine pendulum resulted eventually—and perhaps rather unexpectedly—from a collaboration between Alexander Bruce and the influential Dutch natural philosopher Christiaan Huygens (1629–1695) in November and December 1662.[4,5,6]

**1.2 A career in politics**

It is possible that by 1657 Bruce had originally found himself largely impoverished and exiled on the European continent because of his (British) royalist sympathies and his Episcopalian (Presbyterian) religious leanings, which were at odds with the values of the Puritan Commonwealth established in 1649 by Oliver Cromwell. On the other hand, he may well have moved across the English Channel to boost his family's business interests. In any case, through his commercial contacts and his political leanings, he soon became involved in politics. In 1660 he joined Charles Stuart's itinerant court-in-exile in The Hague, where he became involved in the negotiations leading up to the King's triumphant return on 29 May of that year to England—a period known as the (Stuart) Restoration, in which Charles reclaimed the thrones of England, Scotland, and Ireland for the monarchy after more than a decade of republican rule. Bruce remained in London, '*At the stonecutter's house next to Wallingford House, Charring Crosse*,' in 1660 and 1661.

On 16 June 1659, Bruce had married Veronica van Aerssen, daughter of the wealthy Col. Cornelis van Aerssen van Sommelsdyck, Lord of Sommelsdyck and Spycke, and Louise van Walts, in The Hague. His marriage to Dutch nobility made him a wealthy man, to the tune of 80,000 guilders, thus enabling him to provide for the needs of the exiled King during the monarch's residence in The Hague. Those financial commitments eventually brought Bruce to the brink of bankruptcy, however: by the time of his death, on 9 July 1680, his estate was seriously impoverished by debts owing to the royal cause, prompting the Scottish Court of Session to order a judicial sale.

Meanwhile, King Charles II's Restoration propelled Bruce to a high-profile appointment, in mid-1661, as Privy Councilor of the Scottish government in Edinburgh. Shortly afterwards, in 1663, he succeeded his brother Edward, who never married and hence did not produce an heir, as second Earl of Kincardine. He inherited the family estate at Culross,[7] Fife, where, he returned to his business focus part-time, managing its coal and salt works (Zickermann, 2013), stone and marble quarries, and—through his wife's family—importing luxury articles such as fine furniture and carriages.

In 1667, with Moray and John Hay (1625–1697), first Marquess and second Earl of Tweeddale, he was appointed joint Commissioner of the Scottish Treasury. Their benign administration formed a striking contrast to the oppressive and tyrannical rule of their predecessors. On 10 July 1667, Bruce was appointed an Extraordinary Lord of Session. His political star continued to rise, given that for a while he acted as deputy to the Duke of Lauderdale—Scotland's *de facto* political ruler—at Whitehall Palace in London. His fortunes took a turn for the worse in 1674, however, when he tried to protect the Scottish Presbyterians known as the 'Covenanters,' and went to London to justify to the King his opposition to Lauderdale's policies. Lauderdale had, however, become too influential, and hence an order was obtained for Bruce's removal from public life. With the Earl of Hamilton and a number of other Scottish noblemen, he was—perhaps unfairly—dismissed from the Privy Council in 1676, already disfavoured on the basis of "*the story that has come to the King's ears … of that Communion you were at*."[8] His good friend, Bishop Burnet, seems to suggest that his misfortune may in part have been caused by a measure of gullibility (Burnet, 1724–1734):

> … and it was from him I understood the whole secret of affairs—for he was trusted with everything. He had a wonderful love towards the King, and would never believe me when I warned him what he was to look for if he did not go along with an abject compliance with everything. He found it true in conclusion, and the love he bore the King made his disgrace sink deeper in him than became such a philosopher and so good a Christian.

## 2 SCIENTIFIC PURSUITS

It is in the context of a life spent in business, on managing his estate's affairs, in public service, and in support of the royal cause that we should consider Bruce's remarkable contributions to scientific progress. Although he is credited with making ground-breaking improvements to early pendulum clocks, his scientific and engineering pursuits were most likely accidental, perhaps inspired by a keen nose for business opportunities. After all, Bruce was a businessman first and foremost, and most of his early correspondence with Moray, collected in the *Kincardine Papers,*[9] deals with commercial aspects of coal mining. Nevertheless, this extensive body of correspondence reveals him as a learned, gifted, and deeply religious person. Because of his exile in the Low Countries, his business activities were necessarily somewhat curtailed, he was cut off from his usual social circles, and he had limited hopes of ever returning to his family home—and so he may have found more time to devote to his other interests, including recreational scientific endeavours.

Historical sources disagree, or at least they do not shed unequivocal light onto the question as to whether Bruce had joined a Freemasons' Lodge (Jardine, 2010) in pursuit of his scientific interests. Moray was a Freemason (Stevenson, 1984), and so was Bruce's son, the third Earl of Kincardine. In addition, the elder Bruce must also have known Sir William Davidson (Murdoch & Grosjean, 1995–2020), the Scottish merchant who collaborated closely with the Northern European Masonic networks to restore Charles II to the monarchy. True to his Masonic beliefs, Moray encouraged Bruce, while in exile in Bremen, to continue his studies in Jewish lore, Hermetic chemistry, Paracelsan medicine, Egyptian hieroglyphics, and Masonic symbolism (Schuchard, 2002)—all clearly of interest to Bruce, known as a man

> … of deep personal religion, of highly refined tastes and of very wide attainments; medicine, chemistry, classics, mathematics, mechanical appliances of every kind especially as adapted to

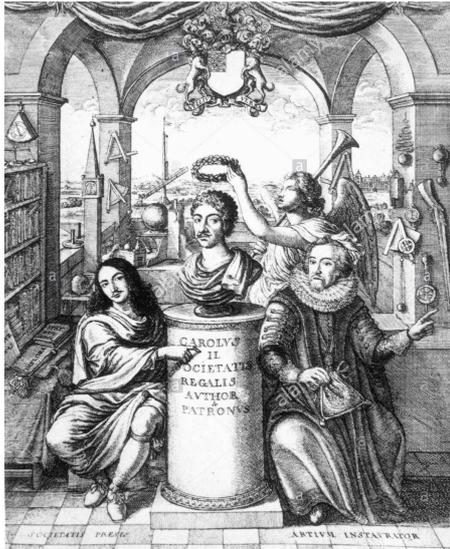

**Figure 1**: Frontispiece of Thomas Spratt's *History of the Royal Society of London* (London, 1667). A garlanded bust of King Charles II stands on a pedestal between William Brouncker and Francis Bacon. (*Public domain*)

his mining enterprises, divinity, heraldry, horticulture, forestry, pisciculture, mining, and the management of estates. (Stephen, 1886)

Whereas Bruce was a member of the Gresham College group of 12 that eventually led to the foundation of the Royal Society of London, his presence at the Society's 28 November 1660 foundation meeting is often attributed to his close friendship with Moray, the Royal Society's *pro tempore* President at the time of its establishment. This notion is supported by the realisation that Bruce did not contribute any meaningful scientific innovations to subsequent Royal Society meetings, despite encouragement from Moray[10] ("*In doing so you will behave your self lyke a true member of the Royal Society, and will be sure to receive their thanks*"; Youngson, 1960) and his appointment, on 30 March 1664, to the Royal Society's Mechanical Committee.[11] Figure 1 is a reproduction of the frontispiece of Thomas Spratt's *History of the Royal Society of London* (London, 1667). It is meant to show the scene at the founding of the Royal Society, including Bruce's triangular Longitude Clock (at top left). Note, however, that while the clock's weight distribution and suspension resemble those of Bruce's novel design, the clock as shown only features a single hand. This may simply be owing to the artist's license (Piggott, 2009b).

Although the Royal Society's records provide evidence that Bruce and Moray were often jointly and highly engaged in Society business (Scala, 1974; Hunter 1989), at least during its early days, Bruce's only contributions to science appear to have been his innovations leading to improved marine timekeepers. He was clearly more interested in furthering his commercial and political interests, but in that sense he was not very different from some of his contemporaries among the Royal Society's founders. The vision originally attributed to Francis Bacon, that material benefits would imminently result from experimental work, was popular among at least a fraction of those early scholars.

**2.1 Marine timekeepers**

Nevertheless, and despite Bruce's low profile at the Royal Society, he was remarkably influential in the early development of marine pendulum innovations. The first record of Bruce's interest in pendulum development dates back to April 1658, when Moray wrote to him (Stevenson, 2007, p. 190; see also Leopold, 1993), still residing in Bremen,

> I haue a second=watch can measure pulses, but no art can make a watch measure 2 minutes equally, unless yong Zulicom [Huygens] at the Hague have found it out, who they say makes clocks that fail not a minute in 6 moneths. But this you will beleave as litle as I do, for I can demonstrate that it must go wrong to keep foot with the sun.[12]

Moray's letter was surprisingly insightful: that final sentence shows that he clearly understood—and was concerned about—what we now know as the 'equation of time,' that is, the need for correction to match clock time to solar time. Just a week later, Moray excitedly followed up and told Bruce, then in Hamburg (Stevenson, 2007, p. 197),

> I have yet to tell you that I have this day seen an exceeding pretty invention of a new way of watch, which indeed I take to be the very exactest that ever was thought upon. … The Rhyngrave [the local Commander in Maastricht, where Moray resided in exile] shew it me. It is long since I heard of it, but did not expect what now I see. The inventor undertakes it shall not vary one minute in 6 moneths, and verily I think he is not much too bold. He is a young gentleman of 22, second son to Zulicon [Constantijn Huygens, who was actually 29 years old at the time], the Prince of Orange's secretary, a rare mathematician, excellent in all the parts of it. I need not describe it to you till we meet, and then I believe I may get you a sight of it.

Apparently, the local Commander had shown Moray his new pendulum clock, since it had developed a defect. Moray's engineering skills were sufficiently fine-tuned that he could identify the problem (Stevenson, 2007, p. 197):

> I find the greatest matter I have at hand to do it with, is that clock I told you of in my last. It is one of the prettiest tricks you ever saw. It stayed no longer here then just to let me see it, as if God had sent it hither of purpose. It was a good part of the time in my hands. It hath a defect and the Rhyngrave sent it to me to consider of, for all that buy them oblige themselves not to put them into workemen's hands. I needed not look upon it long to know all was in it. I needed no more for that than the very first glance I had of it. The rest is but matter of adjusting of numbers for the wheels and pinions.

Although he recommended that the Commander return the clock to its maker, Salomon Coster (1620–1659) of The Hague,[13] he also suggested that Bruce invest in the purchase of one of the new clocks so that they could work on making simple improvements, perhaps appealing to Bruce's business acumen (Stevenson, 2007, p. 199):

> If I thought you had a mind to bestow 40 dollars or some less on one of them I would think to have it ready for you against you come. Never any other design made wanrests [parts of the clock's escapement mechanism] go so equally. … If I make any, I will make it beat another time then this doeth, for it beats at the rate of 80 strokes of the wanrest or thereby to a minute. and I will make it beat just 60 which will be the seconds, and will put an index to shew them. But there is no end of tricks of this kind. When you come to the shop you may perhaps find there will and weal.

Coster's earliest clocks typically included 13.8 cm-long pendulums, and a gear train of 9678 (where the number refers to the number of beats per hour), resulting in approximately 80 'oscillations,' or 2.6 vibrations per second (Whitestone, 2012). As such, these were not typical 'seconds pendulums' used in later years to tackle the longitude problem, and neither were those constructed by most of the early clockmakers following Coster's lead, since seconds pendulums are approximately 99.4 cm long with a 3600 train; a half-seconds pendulum is approximately 24.8 cm long with a 7200 train.

It is clear from this early correspondence that Moray was the driver of Bruce's fledgling interest in the development of a viable marine pendulum. Moray was clearly aware of Huygens' reputation and achievements (Jardine, 2010); the earliest surviving correspondence between both men, dated 22 March 1661, implies that they knew each other quite well already.[14] Meanwhile, Bruce had made Huygens' acquaintance through their shared political views (Jardine, 2010) and their common interest in marine timekeepers. As fate would have it, Bruce had become a frequent visitor to the Van Aerssens, who were Huygens' neighbours, and Bruce soon became a family friend of the celebrated scholar. He was given a presentation copy of Huygens' initial exposition on his first pendulum clock,[15] *Horologium* (1658). At Moray's insistence[16]—who had not yet met the Dutch scholar—Huygens and Bruce engaged in their joint pursuit of a viable marine pendulum from as early as the spring of 1660 (Stevenson, 2007, p. 211):

> If all Mr Zulicom's [Huygens'] addition to his invention be no more but the making of a clock of the size that the pendule beats the seconds, that is every stroak takes up a second, I do not consider that of [importance at] all. For I know the pendule must be about a yard long to do that, and it is believed here that all the church clock's in the Hague are made after his way, so that they ever strike all at once, for so it hath been said here to our queen [Henrietta Maria]. I have not seen his book, nor think it can be bought here. therefore think of sending me one. If you recommend it to Sir Alexander Hume and bid him send it by some of the Earl of St Albans's servants it will come safe. If I see him here I will talk to him of his perspective glasses, and mean to make my court with him upon your account.

**2.2 Collaboration …**

The men became involved in a collaboration that was at times awkward and sometimes downright hostile: Huygens felt that Bruce had elbowed his way into a field that the Dutch scholar had developed. Nevertheless, by the summer of that year, Bruce has ordered, at his own expense, a clock designed by Huygens for Moray, to the latter's great excitement

(Stevenson, 2007, p. 217):

> I am well pleased with Mr Zulicem's ordering of my clock. Let it be so, and I will thank him when I see him. I have not time now to talk of that curiosity you mention, but where people think it needless and that those watches are best that have the pendule fast to the axeltree that hath the two pallets, but I am not yet of their mind, nor for that advantage he speaks of in the stoppers you mention. I shall onely say more of this that if the watch do not mark the inequality of the days, it goes not equally.

Bruce had implemented a number of innovations into the pendulum clock design that preceded Huygens' own efforts. It is unknown whom Bruce consulted for his clock design, but we do know that he moved in the same social circles as Robert Hooke (1635–1703) and Christopher Wren (1632–1723), while he is also believed to have known the London Master clockmakers Ahasuerus Fromanteel (1607–1693) and Edward East (1602–1696). In 1661, when Huygens visited him in London, Bruce demonstrated his new double-fork, 'F'-shaped crutch—equipped with two cranks instead of one—for the first time (see Figure 2; Huygens, 1664), which had been designed to avoid the rigid-body rotation known as 'pendulum banking' allowed by a single crutch. Huygens took a particular interest in Bruce's new design.[17] The known existence of a table clock with an offset winder as fusee—a cone-shaped pulley surrounded by a helical groove, wound with a cord or chain attached to the mainspring barrel—dating from before 1662, fitted with an English dial plate, and signed '*Severyn Oosterwyck Hague*' suggests that Bruce added fusees to his marine pendulum clocks well before Huygens adopted the same approach (Piggott, 2009a). This can be taken as evidence of the London origin of Bruce's original marine timepiece, which he showed to Huygens in 1661. During Huygens' visit to London, he was taken
"*to Fomantils ye famous clock-maker to see some pendules*" (Evelyn, 1661), which suggests that Fromanteel was Bruce's original clockmaker. Huygens' visit to London quickly paid dividends: his friendship with Moray had been cemented, his relationship with Bruce consolidated, and his place among the Fellows of the early Royal Society secured. Perhaps unsurprisingly, Huygens' original goal for the visit, that is, attendance at the coronation of King Charles II, went unfulfilled: he instead prioritised viewing a lunar eclipse that occurred at the same time.

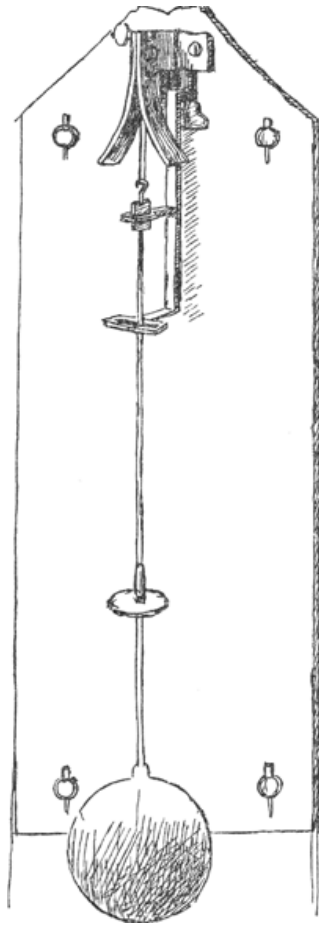

Bruce and Huygens both continued to develop their own marine pendulums. Bruce spent long periods between March and December 1662 in The Hague. In October of that year, on one of his regular return sailings from his Culross estate to the Dutch Republic, he used the journey to test pendulum clocks modified to his own design as potential longitude timekeepers.[18] He commissioned the first Master of the town's Clockmaker's guild, Severijn Oosterwijck (before 1637–c. 1694), to construct two marine clocks to his own innovative, triangular *spring*-driven design for further sea trials in December that year to London (Leopold, 1997, p. 104: note 21).[19] At the time, Huygens was still pursuing rectangular, *weight*-driven designs—or perhaps the Dutch scholar simply depicted his clock design as such in order to confuse and divert competitors. It had become increasingly clear already by the early 1660s—at least to Hooke and Fromanteel, who is thought to have made Bruce's prototype marine pendulum to Hooke's design—that *spring*-driven remontoires were much more practical on choppy seas than their *weight*-driven counterparts. (The gear train of remontoire clocks contains a secondary spring or weight, which drives the clock's escapement; it is rewound regularly by the mainspring or the main weight, thus exerting a steady force on the escapement and ensuring the clock's smooth operation.)

**Figure 2**: Bruce's 'F'-shaped crutch, developed in 1661; drawing from Huygens' 1664 patent application. (*Courtesy Digitale Bibliotheek van de Nederlandse letteren, DBNL*)

Bruce believed that his successful sea trials had convinced Huygens to adopt some of his innovations in his own clock designs. That autumn, the two men spent a lot of time working on their marine timekeeper design together.[20,21] On 4 December 1662, Huygens wrote to his brother Lodewijk that he had been slow in responding to a letter,

> … because of several visits I have received, and principally by that of Mr Brus [Bruce], who did not leave me alone for a single moment all afternoon. And he has been doing that quite often, ever since we set about perfecting our invention for [measuring] Longitudes.

On his return voyage to England in early 1663, one of Bruce's Oosterwijck clocks stopped working (the other apparently remained in The Hague, possibly with Huygens), while his original clock—now known as the 'old' clock (Leopold, 1997, pp. 102–114)—was badly damaged, to the point that it was no longer useful for accurate timekeeping at sea:

> I am glade to find you no more discouraged then my self at the tryell I made of the watches in comeing hither for I will assure you, that no storme lett it be never so violent can so schake a great shippe as the packetboat was in the little wind we hade in comeing over. for you can hardly imagin that any vessell can have so suift a motion as that hade so that the watch which did not fall did schake from syde to syde lyke a pendule it self & with that violence that I wonder it did not fall doune lyke wise. I have not yet gott hither all the peeces belonging to them but I expect them to morow and then I shall show them to Sir Robert Moray & lett yow know their opinions of them. I thinke the best preventive you can use that others do not gett the preference of the invention will be to acquaint your acquaintances of the thing & to desyre that it may not be graunted to any other till the thing be put to a tryell, for to aske any thing till it be perfected I do not so well approve, but I shall submitte to your better judgement & the knowlege you may have of things & persones in that place.[22]

Moray and Huygens continued their exchange of letters following Bruce's seemingly disastrous crossing to England, with both suggesting that the clocks may have fared better had the passage been made on a larger vessel.[23,24] Meanwhile, the London clockmaker John Hilderson (c. 1630–c. 1665) was entrusted with making a copy of Bruce's 'old' clock design, which in turn was taken on Captain Robert Holmes' (c. 1622–1692) voyage to West Africa in 1663–1664. (We will return to Holmes' voyages in Section 3.) By the end of February 1663, Moray wrote to Huygens that he and Bruce were about to embark on additional trials "*at sea, going as far as the Dunes, to try out Mr Bruce's clocks, which he is trying to adjust to the best of his ability*."[25] Note the change in tone at this time, with Moray referring to "*Mr Bruce's clocks*," no longer Huygens'. He continued,

> You are right in saying that the movement of large boats is gentler than that of small ones, but in heavy swells, particularly when the wind is head on, or when the ship is at anchor, the shocks are stronger and more violent. But what I fear most is not the agitation the ship gives to the whole body of the clock (though I am worried that that may have its effect also) but rather that the sudden movements of the ship downwards, and in the contrary direction, which in the one case will make the pendulum slow down, in the other will accelerate it. sometimes making it heavier, sometimes lighter, and either way unequally, which it seems to me is bound to cause deregulation in the mouvement of the clock's mechanism. But it still seems worth testing this experimentally.

Meanwhile, during the summer of 1662, Huygens wrote a number of enthusiastic letters to both his brother Lodewijk, and to Moray, referring to a

> … small pendulum clock ... which works sufficiently well to serve for [the purposes of] longitude determination, and which, once I have given it a push, continues to move without stopping in my room, where it is suspended from 5 foot long ropes, but I have yet to test it on water, for which we should requisition a reasonably sized vessel to [allow us to] sail on choppy seas, something I do not know when I could achieve it.[26]

Upon hearing the news, Constantijn Huygens Sr., their father, appears to have become overly excited, since on 9 November 1662 Huygens asked his brother Lodewijk to urge their father to tone down his enthusiasm:

> I am not as advanced with the invention of [a method to determine] longitudes, as it seems that you believe, and I wish that Father does not talk about it until I have ensured that it is useful. Mr.

Brus [Bruce], who has returned to Scotland, will undertake a sea trial whose success I look forward to, because it is of great importance for these affairs.[27]

Indeed, Huygens was still working on the longitude problem, but meanwhile Bruce had agreed to undertake sea trials, supported by the Royal Society, to test the clocks' reliability and accuracy. Moray, the mathematician William Brouncker (1620–1684), second Viscount Brouncker and first President of the Royal Society, and Hooke (appointed as the Royal Society's 'Curator of Experiments') were all present at Bruce's earliest sea trials (Hooke, 1726; Gunther, 1923–1944). Hooke was clearly impressed:

> The Lord Kincardine did resolve to make some Trial what might be done, by carrying a Pendulum Clock to Sea; for which End, he contrived to make the Watch Part to be moved by a Spring instead of a Weight; and then making the Case of the Clock very heavy with Lead, he suspended it, underneath the Deck of the Ship, by a Ball and Socket of Brass, making the Pendulum but short; namely, to vibrate half Seconds, and that he might be better inabled to judge of the Effect of it, he caused two of the same Kind of Pendulum Clocks to be made, and suspended them both pretty near the middle of the Vessel, underneath the Deck; thus done, having first adjusted them to go equal to one another, and pretty near to the true Time; he caused them first to move parallel to one another, that is, in the Plane of the Length of the Ship, and afterwards he turned one to move in a Plane at Right Angles with the former; and in both these Cases it was found by Trials made at Sea, at which I (i.e. Dr. Hook) was present, that they would vary from one another, though not very much, sometimes one gaining and sometimes the other, and both of them from the true Time, but yet not so much but that we judged that they might be of very good Use at Sea, if some farther Contrivances about them were thought upon, and put in Practice. This first Trial was made in the Year 1662; whereupon, these being found to be able to continue their Motion without stopping, several other Clocks of this Nature were made and sent to Sea, by such as should make farther Experiment of their Use. (Hooke, 1676)

Hooke could not help himself, however, and added that he did not believe that the innovative ball-and-socket ('Cardan') suspension (see Figure 3) would work since they had "*experimentally found [the method of suspension] useless to that effect*" owing to the unpredictable accelerations pendulums were subject to at sea. In March 1663, even Moray had warned Huygens about the inherent flaws of pendulum operation on rocking ships,[28] yet Bruce and Huygens persisted with pendulum-based marine timekeepers.

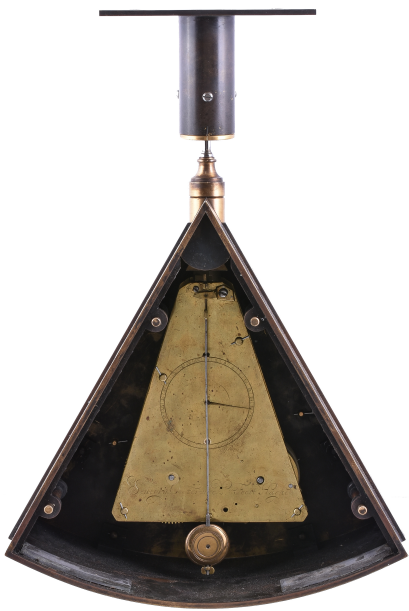

**Figure 3**: One of the few surviving Bruce–Oosterwijck marine timepieces made in 1662. (*Photograph courtesy of Dreweatts 1759*)

Although Hooke indicates that the year of this early sea trial was 1662, this has been called into question. Jardine (2008, 2010) has pointed out that in his Cutlerian lecture of (most likely) 1678, Hooke stated that the trial took place in March 1664, but the actual date may well be closer to March 1663.

**2.3 … and rivalry**

In his letter to Lodewijk of 14 December 1662,[29] Huygens wrote, clearly unhappy, that Bruce claimed intellectual rights to some of the marine clock's design, as well as a share in any profits resulting from its commercialisation, but he did not go into details. Although Huygens eventually but reluctantly agreed to pay him an equal share, Bruce demanded more. He went so far as to dismiss Huygens' contributions as simply an obvious extension of earlier work, which Huygens naturally considered highly insulting. In addition, Huygens felt cornered by the regulation that foreigners could not apply for an English patent,[30] leaving the way open for Bruce to proceed without any regard for Huygens' intellectual contributions. Mediation by Moray on behalf of the Royal Society[31] salvaged the situation to some extent. The result was that Bruce agreed that the patent "*be taken in the name of the [Royal] Society*," which acted both on behalf of Huygens as well as in its own right (Yoder, 1988). While Huygens' perspective on the matter is well-known, Bruce's summary letter of 29 January 1664 is the only surviving document by his own hand that outlines his scientific interests, and so I include it here in full[32] to provide context:

My dear friend                                                                                                        Culross Jan$^r$. 29th—64

     Me thinks I have a great deale more reason to be surprised at what you write to me, than you had to thinke strange of my last to you. For I do very much wonder to find you mention tearmes of agreement, when I am very sure we never entered in treaties. for all that ever we spake of, was only concerning your own contrie, as you may very well remember for I'm sure I do it perfectly. In what concerned that, my respects to you were such that I was willing the patent should pass in your name alone, or in both our names jointly, but as to what concerned other places we never entered in communing about it. And therefor being advertised, of the successe of that tryell which was made of my watches at sea, I would not proceed one steppe in claiming the advantages which may arise from the invention without first acquainting you. And I thought the way I tooke in it was candid enough, that after having told you, what arguments I thought were for my advantage, I added that yet I would be content the differences betwixt us should be determined by discreet persones. And thus I thought to have cut short all further debate. but it seems you are not satisfied with that way since you fall so large upon the debate of it, alledging in the end that all rational men will be of your opinione, which forceth me to say somewhat in this, though I had resolved to argue no more in the matter. And first I shall put you in mynd that at my first arrivall at the Hague, after the tryell I had made betwixt Scotland and that, of my watch, when you did me the favour to see me at my chamber we fell upon the subject of the going of the pendule watches at sea, and you told me positively then that it was your opinione that it was impossible, that you had been making experiments of it, and all the effect of them was, to be settled in that opinion by them. You did lykewise urge reasons of the impossibility of it. I came afterwards to see that watch by which you had made your experiment and I believe you will acknowledge that it was so farre different from mine in the whole way of it that it is not lyke they should ever have met. And the rather I think this that I showed you at London 18 months before that tyme the same watch which received very small amendments thereafter, and if you had thought that way able to bring it to passe, you might from that view have ordered one to be made for your tryell. And now after you hade given over all hopes of it your self, even of this same way too, you having seen it before quiting of your hopes. When I come to make a tryell of it and find it answer my expectations as much as the disorder the watch was then in, could allow: and then out of friendshippe, trust the discovery of it to you, though I confesse not under any ingadgement; thinking it below the friendship I bore you, to exact any of you, that now I say, you will pretend further in that invention, than what I would allow, I think very strange. There is no body can denye but that the first application of a pendule to a watch is yours, but I think that you can as little denye that without this additione of mine (as little subtile as it is) that could not have been usefull at sea. And though indeed I do not at all think it subtile yet I am sure it is my inventione, and it is not always the subtilest inventiones that are the most usefull. for truly I do not looke upon the application of the pendule to a watch as at all subtile since I am very confident that I know many a man that if it had but entered in their thoughts to applye the one to the other they would easily and quickley have done it, and yet I thinke it an excellent inventione and very usefull. And upon the whole matter I am confident that any rationall man will judge that no other persone should pretend the advantages of this inventione, but he that brought it to the desyred end, all the previous inventions which are made use of in it, being already published to the world, for showing what cloke it was. And therefore I shall leave you yet to your further thoughts, and waite for the knowledge of them, for I shall be very unwilling to have difference with a persone for whom I have so great a kyndness, especialy in so emptie a debate as this may yet chance prove for all the fine appearances that are yet.

I am My deare friend Your most affectionat friend & servant
          A. Kincardin

This version, that is, Bruce's version of events, is confirmed in the transcript of Hooke's 1674/5 Cutlerian lecture (Hooke, 1679). The 29 January 1664 letter is included among the *Kincardine Papers*, but it actually never reached Huygens (Jardine, 2010). It was intercepted by Moray, who must have considered it more prudent to withhold the letter so as not to jeopardise his mediation efforts.

Publicly, Huygens' attitude seemed to have consisted of simply ignoring Bruce's claims: even when Bruce and the Royal Society were granted an English patent on 13 March 1665,

Huygens deferred any discussion of their clock's design or Bruce's contribution,[33] of which the most important aspects were the use of a steel ball encased in a brass cylinder as the pendulum's pivot[34] and the application of the 'F'-shaped double crutch to keep the pendulum swinging in a single plane—which he introduced in addition to the application of a stabilising Cardan suspension. Both innovations were included in Huygens' patent drawings of November 1664,[35] although the Dutch scholar appears to have adopted the attitude that Bruce's improvements did not merit naming him as co-inventor. We have to wait until his publication of *Horologium Oscillatorium* in 1673 to unveil some of the mystery surrounding the Huygens–Bruce marine clock design (Huygens, 1673, pp 16–17; Mahoney, 1980):

Instead of a weight they had a steel strip wound in a spiral, but the force of which the wheels were turned 'round, just as is commonly employed in those small watches that are wont to be carried about. So that the clocks could endure the tossing of the ship, he [Bruce] suspended them from a steel ball enclosed in a brass cylinder, and extending downward the arm of the crutch that sustains the pendulum's motion (the pendulum, by the way, was a half-foot in length) he doubled it to resemble the form of an inverted letter F; namely, lest the pendulum's motion wander out in a circle with the danger of stoppage. (see also Figure 4; Huygens, 1664–1665)

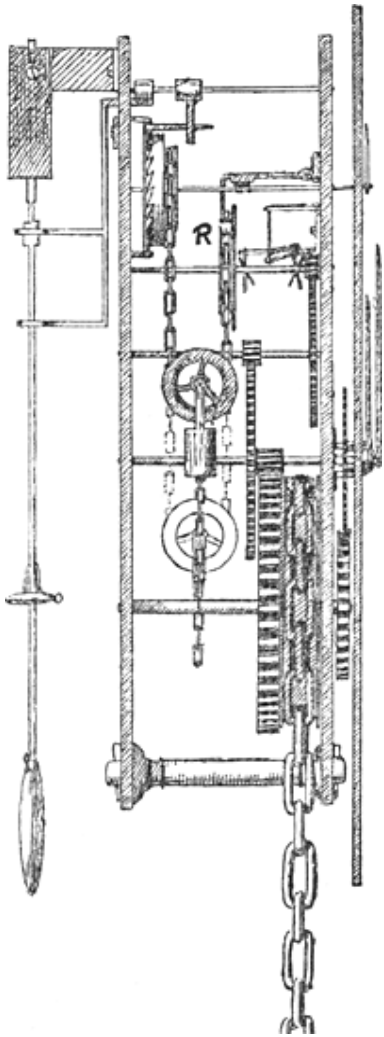

**Figure 4**: One of many examples of Huygens' sketches in support of the development of his pendulum clocks in 1663–1665. Note the 'F'-shaped crutch. (*Courtesy Leiden University*)

Moreover, Bruce preferred clocks with short pendulums for improved portability. In addition, he designed methods of support and suspension he hoped would protect the clocks from the violent pitching and rolling of ships tossed about by storms and high seas. His innovations were marked improvements over Huygens' simple rope suspensions (Leopold, 1993). Yet, Huygens could not bring himself to name Bruce as co-inventor; he only referred to him as "*a Scottish gentleman and a friend of ours*" (Huygens, 1673, p. 16). Huygens also attempted to alter the historical record by asserting that the first British sea trial of a marine pendulum had occurred in 1664, although that is the year that Huygens himself got involved; Bruce's earliest sea trials date back to at least 1662.

**3 CAPTAIN HOLMES' SEA TRIALS**

Huygens had already considered a number of different ways in which to reduce friction and suspend a clock from a pivot so that the ship's motion would not affect the clock's operation (Mahoney, 1980). Bruce's idea of using a steel ball inside a brass cylinder turned out to be highly effective and stable in tests performed in The Hague. Huygens shipped two clocks featuring this new design to Bruce at the Royal Society in London in early 1663. They were eventually sent on sea trials to Lisbon, Guinea, and into the Atlantic Ocean, under the command of Holmes on board the English Royal Navy frigate *H.M.S. Reserve*.[36] The first results were obtained between 28 April and 4 September 1663. They exceeded expectations, confirming that both clocks ran highly reliably, to within a few minutes over 24-hour time intervals, and allowed longitude determinations which were within a few minutes of arc of those determined by independent means in well-known waters. Holmes' favourable report of the clocks' operation[37] further strengthened Bruce in his belief that he should be given a patent for 'his' clock design (Leopold, 1997, pp. 102–114; Mahoney, 1980, notes 53–54). However, Samuel Pepys (1633–1703), Clerk of the Acts to the Navy Board, politician, and well-known diarist, questioned the accuracy of Holmes' measurements—and indeed, Holmes had reported implausible success, beyond even the Royal Society's best hopes (Jardine, 2008, Ch. 10):

> The said master [the captain of Holmes' ship] affirmed, that the vulgar reckoning proved as near as that of the watches, which, added he, had varied from one another unequally, sometimes backward, sometimes forward, to 4, 6, 7, 3, 5 minutes; as also that they had been corrected by the usual account.

Nevertheless, given their early success, Moray and Brouncker offered their backing to Huygens and Bruce to market their clocks in the Dutch Republic, France, Spain, Sweden, and Denmark. They had previously attempted to patent Hooke's spring-driven pendulum clock, but Hooke had aborted those efforts in 1660 because of the restrictive terms associated with the proposed application (de Grijs, 2017, p. 4-2). Their new efforts to turn the Huygens–Bruce clocks into commercial success did not work out either (Patterson, 1952).

Holmes' second voyage, in 1664 on the *H.M.S. Jersey*, proved to be a crucial and dramatic confirmation of the promise of the Huygens–Bruce marine clock design. The captain had set sail due west from St. Thomas off the coast of Guinea, heading out some 800 leagues (approximately 4500 km) into the Atlantic Ocean before revising their course northeastwards to the coast of Africa. When, after a few days the ship's stores and, particularly, their fresh water supply began to dwindle, Holmes' senior crew recommended that they reroute towards Barbados. Their traditional piloting methods placed them some 100 leagues from Cape Verde—a 3–4 day run—but longitude determinations based on Holmes' clock suggested that they were only some 30 leagues out of port. Holmes decided to trust his own measurements and continue their predetermined course; they reached the islands the next afternoon.[38] The Huygens–Bruce clock had thus proved its potential as a maritime navigation aid; the traditional method had been unable to detect that the ship had drifted with the ocean currents some 80 leagues eastwards.

Obtaining details about the voyage after its completion turned out to be troublesome, however: upon his return to English shores, Holmes was immediately incarcerated in the Tower of London, allegedly for unwarranted hostile actions against foreign interests.[39] Indeed, although Holmes' orders, signed by King James II of England, were to "*promote the Interests of the Royall Company*"—that is, the Royal African Company, which had been established to trade along the west coast of Africa—and to "*kill, take, sink, or destroy such as shall oppose you*", he had been too successful; more successful even than anticipated by the most unreasonable (that is, commercial and greedy) expectations.

James's predecessor, King Charles II, had expressly promoted pursuing economic dominance, to the detriment of the Dutch. He clearly hoped that a combination of English naval force and state-sanctioned piracy would cripple the Dutch Republic financially and force its Staten Generaal (the Republic's government) to agree to a favourable peace settlement. Instead, Holmes' actions, which consisted of having captured Dutch forts on the West African coast as well as half a dozen ships, led to the Second Anglo–Dutch war (1665–1667). This had significant repercussions back in London, and so Holmes was made a scapegoat and charged with exceeding his orders. However, it was not his military prowess that landed him in the Tower of London, but the greed of the Royal African Company's governors—Holmes' takings in merchandise were far behind the Company's rather unrealistic, materialistic expectations.

His imprisonment was cut short, however, by the Dutch declaration of 22 February 1665—which was interpreted by the English as a declaration of war—that they would retaliate against British shipping. Holmes' leadership and naval expertise were sorely needed again. Meanwhile, by early March 1665 Moray, acting on Huygens' behalf, tracked down another of the *H.M.S. Jersey*'s officers, who provided enough insights to suggest that the original data needed updating.[40] Yet, Moray kept the information about the ship having drifted on account of the ocean currents to himself until 27 March 1665.[41]

Huygens, nevertheless enthralled by the success of this latest voyage, proceeded apace. He published an account of the voyage and its scientific success in the 23 February 1665 issue of the *Journal des Sçavans* and made his clock available for public sale. He offered anyone interested in purchasing his clocks detailed instructions regarding their regulation and use, collected in his *Kort Onderwys aengaende het gebruyck der horologiën* *tot het vinden der lenghten van Oost en West* (1665),[42] which was translated into English in 1669, titled *Instructions Concerning the Use of Pendulum-Watches, for find the Longitude at Sea*. It contained the first known table of the equation of time—although Huygens used conventions that differ from those commonly applied today: he computed the equation of time as a positive number, setting his zero-point value at 11 February. To reconcile his values with the position of the 'mean' Sun, one should subtract the annual average, 14 min 25 s, from his values (Grimbergen, 2004).

On 16 December 1664, the Staten van Hollant en Westfrieslant (the Provincial States of Holland) awarded Huygens a 15-year patent for his marine clock.[43] A series of letters to Moray at the Royal Society and Jean Chapelain (1595–1674) at the *Académie Française* in the spring of 1665 show that this prompted Huygens to seek the same recognition from the French and English governments.[44] While his patent applications were being considered, he halted his plans to publish his new treatise, *Horologium Oscillatorium*, which contained full details of his clock's design and which he had been working on since making the first revisions to its precursor, *Horologium*, in 1660. Clearly, if he wanted to make a profit from his work, particularly now that his first marine clock looked like it might sustain rigorous sea trials and form the core of a practical and, most importantly, profitable method of longitude determination, he should not publicly release the clock's design details so as not to jeopardize the exclusive license he was after. In addition, he expected to soon be able to add more detailed and systematic data in support of the clock's seaworthiness (Mahoney, 1980).

Yet, despite the concerted efforts of 17th-century heavyweights like Bruce, Huygens, Moray, Hooke, and their contemporaries, pendulum clocks never became viable marine timekeepers. Eventually, spring-driven watches took centre stage, although we would need to wait another century (de Grijs, 2017)[45] before the unassuming English clockmaker John Harrison (1693–1776) managed to find a workable solution to solving the perennial longitude problem through application of innovative metallurgical techniques. Nevertheless, Bruce's innovations stood the test of time and so his little-known Scottish inventions have since been widely incorporated in subsequent designs of generations of pendulum clocks.

## 4 NOTES

[1] The royalists spent significant efforts in courting the Swedes, not in the last place because of the presence of (and possible support from) a sizable Scottish community in Gothenburg (e.g., *Kincardine Papers* MS 5050, f. 28, 18 April 1658).

[2] Sir Robert Moray seems to have been a confidante and mentor of both Bruce cousins. Some sources suggest that Alexander Bruce was Moray's protégé. Cruickshank (2012) interprets letters from Moray, addressed to "*Will*," as affectionate (see also Youngson, 1960, p. 255; Keblusek 2004).

[3] 1657-10-06: Moray, Robert – Bruce, Alexander; *Kincardine Papers*, ff. 7–8.

[4] 1662-12-01: Huygens, Christiaan – Moray, Robert; *Oeuvres Complètes de Christiaan Huygens*, IV, 274–275 (No. 1080).

[5] 1662-12-20: Huygens, Christiaan – Moray, Robert; *Ibid.*, IV, 280–281 (No. 1083).

[6] 1662-12-14: Huygens, Christiaan – Huygens, Lodewijk; *Ibid.*, IV, 278 (No. 1082).

[7] Royal Commission on Ancient and Historical Monuments and Constructions of Scotland: Fife, Kinross and Clackmannan, 1933. Edinburgh: HMSO, 69–87.

[8] 1665-08-15: Moray, Robert – Bruce, Alexander; *Kincardine Papers*, ff. 202–203.

[9] The *Kincardine Papers* (14 September 1657–2 June 1673) represent the correspondence between Moray and Bruce, collected in more than 120 letters.

[10] 1662-10-27: Moray, Robert – Bruce, Alexander; *Kincardine Papers*, ff. 164–165.

[11] Youngson (1960) suggests that the Earl of Elgin included among the committee's membership (who was not a Fellow of the Royal Society) was likely meant to refer to the Earl of Kincardine.

[12] Moray's handwriting was appalling and his grammar and spelling were often inconsistent (e.g., Stevenson, 2007). This is reflected in some of the passages included in this article.

[13] Since Huygens' first clock in 1657, Coster signed his clocks "*Samuel Coster Haghe met privilege*," as indication that he was the authorised clock maker.

[14] 1661-03-22: Moray, Robert – Huygens, Christiaan; *Oeuvres Complètes de Christiaan Huygens*, III, 260–261 (No. 851).

[15] 1658-09-06: Huygens, Christiaan – de Sluse, René-François; *Ibid.*, II, 209–210 (No. 503), note 2.
[16] 1662-12-01: Huygens, Christiaan – Moray, Robert; *Ibid.*, IV, 274–276 (No. 1080).
[17] 1664-01-29: Bruce, Alexander – Huygens, Christiaan; *Ibid.*, XXII, 605–609.
[18] 1662-11-09: Huygens, Christiaan – Huygens, Lodewijk; *Ibid.*, IV, 256–257 (No. 1073).
[19] A surviving copy has recently been found and is on display at the National Museum of Scotland: https://www.nms.ac.uk/explore-our-collections/stories/science-and-technology/bruce-oosterwijck-sea-clock/ [accessed 14 Jan 2020].
[20] 1662-12-14: Huygens, Christiaan – Huygens, Lodewijk; *Oeuvres Complètes de Christiaan Huygens*, IV, 278–280 (No. 1082).
[21] 1662-12-28: Huygens, Christiaan – Huygens, Lodewijk; *Ibid.*, IV, 284–285 (No. 1086).
[22] 1663-01-16: Bruce, Alexander – Huygens, Christiaan; *Ibid.*, IV, 301–302 (No. 1095).
[23] 1663-01-19: Moray, Robert – Huygens, Christiaan; *Ibid.*, IV, 295–299 (No. 1093).
[24] 1663-02-20: Huygens, Christiaan – Moray, Robert; *Ibid.*, IV, 304–306 (No. 1097).
[25] 1663-03-01: Moray, Robert – Huygens, Christiaan; *Ibid.*, IV, 318–321 (No. 1102).
[26] 1662-06-09: Huygens, Christiaan – Moray, Robert; *Ibid.*, IV, 148–152 (No. 1022).
[27] 1662-11-09: Huygens, Christiaan – Huygens, Lodewijk; *op. cit.*
[28] 1663-03-01: Moray, Robert – Huygens, Christiaan; *op. cit.*
[29] 1662-12-14: Huygens, Christiaan – Huygens, Lodewijk; *op. cit.*
[30] 1664-09-23: Moray, Robert – Huygens, Christiaan; *Ibid.*, V, 115–117 (No. 1256).
[31] 1664-01-09: Huygens, Christiaan – Moray, Robert; *Ibid.*, V, 6–7 (No. 1200).
[32] 1664-01-29: Bruce, Alexander – Huygens, Christiaan; *Ibid.*, XXII, 605–606.
[33] 1663-12-09: Huygens, Christiaan – Moray, Robert; *Ibid.*, IV, 1458–1461 (No. 1178); see also Huygens' correspondence in the spring of 1664; *Ibid.*, V.
[34] 1663-01-02: Bruce, Alexander – Huygens, Christiaan; *Ibid.*, IV, 290–291 (No. 1090).
[35] 1664-11: Huygens, Christiaan – Staten Generaal; *Ibid.*, V, 152–154 (No. 1278).
[36] The first voyage took place from 29 April to 4 September 1663: *Ibid.*, IV, 446–451 (No. 1174; appendix to correspondence between Holmes and Moray); Huygens' response is contained in correspondence between 29 October and 9 December 1663: *Ibid.*, IV, 426–474.
[37] See also the *Lauderdale Papers*, Nat'l Library of Scotland, MS 13500, ff. 35–36.
[38] Holmes reported the 1664 Atlantic voyage to Moray; see 1665-01-23: Moray, Robert – Huygens, Christiaan, *Oeuvres Complètes de Christiaan Huygens*, V, 204–206 (No. 1315); see also Holmes (1665–1666).
[39] 1665-02-13: Moray, Robert – Huygens, Christiaan: *Ibid.*, V, 233–238 (No. 1329).
[40] 1665-03-13: Moray, Robert – Huygens, Christiaan: *Ibid.*, V, 268–272 (No. 1353).
[41] 1665-03-27: Moray, Robert – Huygens, Christiaan: *Ibid.*, V, 284–288 (No. 1363).
[42] Huygens (1665); 1665-01-02: Huygens, Christiaan – Moray, Robert: *Ibid.*, V, 185–189 (No. 1301). Huygens engaged a printer on 19 February 1655 and sent the page proofs to Moray on 27 February 1665 so as to facilitate an English translation. The latter was published in *Phil. Trans. R. Soc.*, 4, 937–953 (1669). A French translation was postponed until more data had been acquired.
[43] 1664-16-12: Staten van Holland – Huygens, Christiaan; *Oeuvres Complètes de Christiaan Huygens*, V, 166–167 (No. 1286).
[44] Huygens corresponded extensively with both Moray and Chapelain in the spring of 1665; see *Ibid.*, V.